\documentstyle[11pt,a4,epsf]{article}
\renewcommand
\baselinestretch{2.0}

\renewcommand{\baselinestretch}{1.25}
\def\beql[#1]         {\begin{equation} \label{#1} }
\def\beq              {\begin{equation} }
\def\eeq              {\end{equation}   }
\def\beqal[#1]        {\begin{eqnarray} \label{#1} }
\def\beqa             {\begin{eqnarray} }
\def\eeqa             {\end{eqnarray}   }
\def\eeqann           {\nnum \end{eqnarray}}
\def\nnum             {\nonumber}

\def\scap             {\noindent \ul}
\def\parag            {\vspace{2ex}}
\def\ul               {\underline }     

\def\r[#1]            { \quad {\rm #1} \quad }

\def\kd[#1]           {\delta_{#1}}     
\def\d,#1             {\partial_{#1}}   

\def\1,#1             {{1 \over {#1}}}
\def\ma[#1,#2,#3,#4]  {{\left( \matrix{ #1  & #2 \cr
                                        #3  & #4 \cr } \right)}}
\def\ve[#1,#2]        {\left( { #1 \atop #2 } \right) }
\def\bra[#1]          {{\big\langle} {#1} {\big|}}
\def\ket[#1]          {{\big|} {#1} {\big\rangle}}
\def\bk[#1,#2]        {{\big\langle} {#1} {\big|} {#2} {\big\rangle}}
\def\ev[#1]           {{\big\langle} {#1} {\big\rangle}}
\def\com[#1,#2]       {{\Big[} {#1} , {#2} {\Big]}}
\def\acom[#1,#2]      {{\Big\{} {#1} , {#2} {\Big\}}}

\def\ferMa   {m_{\it f}}                        
\def\cc      {g}                                

\def\lcx     {x^-}                              
\def\lctim[#1] {{#1}^+}                         
\def\lcspa[#1] {{#1}^-}                         
\def\zm[#1]  {\langle {{#1}} \rangle _0}        
\def\nm[#1]  {\langle {{#1}} \rangle _n}        
\def\lcHam   {H_{\rm LC}}                       
\def\dsfr    {\Psi _R }                         
\def\dsfl    {\Psi _L }                         

\def\kinE             {T}                       
\def\resset           {{\cal D}}

\def\massev2          {M^2}
\def\mass2            {M^2}
\def\res              {K}
\def\contractions     {C}
\def\seagulls         {S}
\def\forks            {F}
\def\qb               {\bar{q}}

\def\Cq0              {C_{q \to q}^{(0)}        (1)}
\def\Cqb0             {C_{\qb \to \qb }^{(0)}   (1)}

\def\boxLe            {L}

\def\emPot            {A}

\def\momentum         {P^+}
\def\energy           {P^-}

\def\emo              {k}
\def\pmo              {p}
\def\enu              {N}

\def\change[#1,#2]    {c_{{#1},{#2}}}
\def\rsign            {\Sigma_r}
\def\spect            {{\Delta_s}}

\def\conjMom[#1]      {\Pi_{#1}}

\begin{document}
\title{\bf
Discretized Light-Cone Quantization:
Solving a 1+1 dimensional field theory (QED)
}
\author{
  Stephan Elser\thanks{Present Address:
  Institut f\"ur Physik/COM,
  Humboldt-Universit\"at, Invalidenstrasse 110,
  D-10115 Berlin, Germany}, Hans-Christian Pauli {\rm{and}} Alex C. Kalloniatis
\\
  {\it Max-Planck-Institut f\"ur Kernphysik} \\
  {\it Postfach 10 39 80 }\\
  {\it D-69029 Heidelberg, Germany} }

\maketitle

\vspace{-0.9truecm}
\begin{center}
Preprint MPIH-V17-1995
\end{center}
\vspace{0.3truecm}

\begin{abstract}
We describe the programming method for generating the
spectrum of bound states for relativistic quantum
field theories
using the nonperturbative Hamiltonian approach of Discretized
Light-Cone Quantization. The method is intended for
eventual application to quantum chromodynamics in
3+1 dimensions. Here the fundamental principles are
illustrated concretely by treatment of QED in two dimensions.
The code is intended as a basis for extensions to
include more complicated gauge symmetry groups, such
as SU(3) color, and other quantum numbers.
The code was written in Fortran 77 and implemented on
a DEC 5000-260 workstation with a typical runtime of
0.2s at total momentum 10.
\end{abstract}

\vspace{0.4cm}


\vspace{-0.1cm}
{\parindent 0pt
{\Large \bf Program Summary } \\[0.0cm]

{\it Title of program:} ms\_main.f \\[0.15cm]
{\it Catalogue number:} ???? \\[0.1cm]
{\it Program obtainable from:}
MPI f\"ur Kernphysik, Heidelberg, FRG \\[0.15cm]
{\it Licensing provisions:} none \\[0.15cm]
{\it Computer:} workstation DEC 5000-260;
{\it Installation:} MPI f\"ur Kernphysik, Heidelberg, FRG\\[0.15cm]
{\it Operating System:} ULTRIX version 4.2 \\[0.15cm]
{\it Programming language used:} Fortran 77 \\[0.15cm]
{\it Program storage required:} 255kbyte \\[0.15cm]
{\it Data storage required:} 80kbyte \\[0.15cm]
{\it Peripherals used:} terminal \\[0.15cm]
{\it No. of bits per word:} 32 \\[0.15cm]
{\it Number of lines in program:} 1894 \\[0.15cm]
{\it Nature of the physical problem} \\
Gauge field theories are notoriously difficult to
solve non-perturbatively, in particular non-Abelian
gauge theory which underlies quantum chromodynamics.
We describe the relatively new approach of [1] to this problem,
Discretized Light-Cone Quantization (DLCQ), which
in many respects is complementary to Lattice Gauge Theory.
We detail the method in some generality but present
the programming application to Quantum Electrodynamics in 1+1
dimensions [2].
Mathematically, in DLCQ one formulates the theory
in light-cone coordinates on a finite space-interval or
volume with periodic (and/or antiperiodic) boundary conditions on
the quantum fields. Correspondingly the plane wave Fock basis
is discretized. A particle number and momentum cut-off lead to
a completely specified numerical problem. One can thus solve the
Schr\"odinger equation for bound states.
The continuum limit must be obtained by numerical extrapolation of,
for example, the invariant masses.
The generalization to other theories is evident. The program
is intended as a template for further
calculations in theories approaching the complexity of QCD(3+1),
and has been already used as such.
\\[0.25cm]
{\it Method of solution} \\
The algorithm can be briefly described as follows:
\begin{description}
\item[{Step 1:}]
Choose a harmonic resolution and particle number cut-off.
\vspace*{-0.3truecm}
\item[{Step 2:}]
Construct the Hilbert space (Fock states)
respecting statistics and symmetries.
\vspace*{-0.3truecm}
\item[{Step 3:}]
Evaluate for each matrix element the
analytic expression, making sure all conser\-vation laws
are obeyed.
\vspace*{-0.3truecm}
\item[{Step 4:}]
Add all allowed terms to obtain the Hamiltonian matrix.
\vspace*{-0.3truecm}
\item[{Step 5:}]
Diagon\-alize, yielding the complete mass spectrum and
wave functions.
\vspace*{-0.3truecm}
\item[{Step 6:}]
Repeat steps 1-5 for higher resolutions to check
convergence.
\end{description}
{\it Restrictions:}
The resolution is restricted by
the available memory, as the Hamiltonian matrix
is stored. The run time is negligibly
restrictive
for matrices of dimension 1500,
but grows with matrix dimension $d^3$.\\[0.25cm]
{\it Typical run time:}
0.2s for the test case of resolution 10. \\[0.25cm]
{\it References:} \\
$[$1$]$ H.C. Pauli, S.J. Brodsky, Phys. Rev. {\bf D32} (1985) 1993, 2001 \\
$[$2$]$ J. Schwinger, Phys. Rev. {\bf 128} (1962) 2425 \, \,
}

\newpage
{\centering \Large \bf Long Write-Up}

\section{Introduction}
The major physical theory calling for the development
of {\it nonperturbative} techniques for solving quantum
field theories is quantum chromodynamics (QCD). It is arguably
{\it the} theory of strong interactions describing hadrons.
However to extract the physics of mesons and baryons
within a picture of relativistic bound states of quarks
and gluons is extremely difficult. Lattice Gauge Theory
\cite{kogut79}
is one nonperturbative method which is presently computing ratios
of hadron masses with ever improving success \cite{butler93}.
However it offers little insight into how the above constituent
or {\it partonic} picture emerges out of QCD.
Discretized light-cone quantization (DLCQ) exploits the
computational elegance and simplicity of the lattice method --
by working with a {\it momentum space} lattice. However it
also employs the physically intuitive power of Dirac's
`front form' \cite{dirac49} in which the vacuum is essentially trivial and
one of the momentum operators is independent of the interaction
and positive definite. Thus
the low energy part of the QCD hadron spectrum
in such a framework would be well described by a
low number of light-cone Fock space quanta built from the simple
light-cone vacuum. DLCQ obtains both the spectrum and the
wavefunctions of relativistic bound states of the theory by
direct diagonalization on the computer of a finite dimensional
Hamiltonian matrix. How a Lagrangian field theory is reduced
to such a problem is the whole art of DLCQ.

This is not the place to detail the history or subtleties of
DLCQ for which the reader is referred to \cite{overview}.
It suffices to say that QCD in (3+1) dimensions
has not yet been attacked
with the full power of the method. Rather it is being
worked towards via progressively more complicated
models by ourselves \cite{woelz} and others
\cite{hiller93,vandesande94,demeterfi94,tam94}.
This is therefore an opportune
point in the development of the method to present the
programming details at their simplest but with enough richness
to display the power of the method.
We turn to quantum electrodynamics in one space and
one time dimension, QED(1+1), originally treated by
Eller \cite{epb}. This model shares with QCD(3+1) the feature of
being a relativistic theory of gauge vector bosons coupled to
fermions. It is, however, unencumbered by the
technicalities of color and transverse dimensions. These
demand a richer Fock space and additional Hamiltonian structures
the programming of which
can nevertheless be grafted onto the basic procedures we outline.
Moreover the simplicity of the theory here means Hamiltonian
matrix elements can be analytically calculated by hand. This may be
technically inefficient in more elaborate theories in which case part of this
task could also be programmed.
Two aspects, however, could change the
underlying method of programming: nontrivial renormalization
and dynamical zero modes of gauge boson fields. These problems
certainly become acute in higher dimensions and remain
at the frontier of present research into the method
\cite{perry,kalloniatis} but we will
not say anything more about them in the present work.

In the following section we outline briefly how one builds
from the Lagrangian defining the field theory to the
matrix problem via DLCQ. In Section 3 there is a description of
the various components of the program and how they can be implemented
in a test run. The final section gives a statement of conclusions.


\section{Light-Cone Hamiltonian Matrix}
\label{hamiltonian}
This section is intended to explain notation and concepts
in the formulae and the corresponding computer codes.
To show the generality of the method we describe the
procedure for QCD(3+1), and later show how this concretely
is worked out for the model case QED(1+1).

\noindent
\scap{Derivation.} For a given quantum field theory in
an arbitrary number of space-dimensions,
DLCQ sets out to solve the eigenvalue equation
\beqa
\lcHam \ket[\Psi ] = m^2 \ket[\Psi ] .
\label{schrodeq}
\eeqa
Here $\lcHam \equiv P^\mu P_\mu =  2 \energy \momentum - P_\perp^2$
is the so-called `light-cone Hamiltonian'. In two dimensions $P_\perp = 0$.
Evidently Eq.(\ref{schrodeq}) is a realization of the
relativistic energy-momentum relationship on physical eigenstates of
the given theory. The (four) momentum operator $P^\mu$ is built from
the energy-momentum tensor, a conserved quantity under evolution in
light-cone time
$x^+ \equiv \frac{1}{\sqrt{2}}(x^0 + x^3)$.
Our conventions follow Kogut and Soper \cite{kogut70}.
Initial data for the independent Lagrangian fields are
specified on the null-plane surface $x^+=0$. This will be done
by expanding the {\it independent} fields at $x^+=0$ in momentum modes.
A momentum space `lattice' is obtained
by restricting the one
dimensional light-cone space variable
$\lcx \equiv \frac{1}{\sqrt{2}}(x^0 - x^3)$ to a finite interval
from $-\boxLe $ to $+\boxLe $ and invoking boundary conditions on the fields.
A similar demand can be made on $x_{\perp}$ in higher dimensions giving
discretized $k_\perp$.
Periodic boundary conditions for bosons is the only choice that
is consistent with the standard Euler-Lagrange equations.
As fermions appear bilinearly in the Lagrangian of any of
our candidate gauge theories, either periodic
or antiperiodic boundary conditions can be used.
The program we present in the sequel allows the
user to choose. Thus the corresponding momenta $k^+$
are discretized in, respectively, integer or half-integer units of
${\pi/L}$. Due to light-cone kinematics, these integers are either
positive or zero. Ignoring zero modes of gauge bosons then,
we can choose the light-cone gauge $\emPot ^+=0$.
The Maxwell and Dirac equations give constraints allowing one to
eliminate the gauge potential $\emPot ^-$ and the lower spinor component $\dsfl
$.
Thus, one is brought to the independent degrees of freedom, which
in physical dimensions are the transverse gluons $\emPot ^i, i = 1,2$
and the upper spinor component $\dsfr $. These are subject to quantization.
Quantization is achieved by imposing
equal $x^+$ (anti)commutation relations on the dynamical fields.
These translate into (anti)commutators
on the Fourier coefficients of the aforementioned momentum mode expansions.
The coefficients become Fock operators of boson, fermion and antifermion
states.
The above eigenvalue problem is then expressed in terms of these
operators: the light-cone Hamiltonian is written in second quantized
form, and the wavefunctions are built from the Fock operators as an
expansion in the number of fundamental partons with amplitude coefficients
which are obtained by solving Eq.(\ref{schrodeq}). The eigenvalue
problem for the bound states is thus completely specified numerically.

Now we specify using the model theory QED(1+1).
First, in 2 dimensions there is no spin and,
neglecting zero modes, there are no dynamical photons. Thus
$\dsfr $ is the only independent quantum field leading to fermion and
antifermion Fock operators in terms of which the eigenvalue problem must
be specified.
In the light-cone Hamiltonian one can separate out the dimensionful
Lagrangian parameters of fermion mass $\ferMa $ and coupling constant $\cc $
by invoking the dimensionless operator functions in
$\res ,\kinE, \contractions ,\forks $ and $\seagulls $
\beqa
\lcHam = 2 \res (\ferMa ^2 + \frac{\cc ^2}{\pi})
\Bigl[
(1- \frac{1}{1+\pi(\frac{\ferMa }{\cc } )^2 } )
\kinE
+ \frac{1}{1+\pi(\frac{\ferMa }{\cc } )^2}
(\contractions + \forks + \seagulls )
\Bigr] .
\eeqa
These operators are built from combinations of the fermion and
antifermion Fock operators and analytic expressions for their
matrix elements will be given in the sequel.
Here $\res $ represents the dimensionless
momentum called the harmonic resolution
and $\kinE $ is the kinetic energy. In the literature,
the potential energy terms
$\contractions $, $\forks $ and $\seagulls $
are called respectively `contractions', `forks' and `seagulls'
because of a diagrammatic representation of their
operator structure
\cite{pauli85,elser94}.
The simplicity of QED(1+1) allows application of the Wick expansion
leading to analytic expressions for the
matrix elements of $\lcHam $ between arbitrary Fock states.
These expressions are given below. The matrix is normalised
by a factor $(\ferMa ^2 + \frac{\cc ^2}{\pi } )^{-1}$.

\parag
\scap{Notation.} In the absence of spin and color, the
only quantum number needed to specify states is longitudinal
momentum. In the following,
we use
$\emo $ to represent this momentum for single electron states
and $\pmo $ for positron states.
The quantity $\enu $ will represent the particle number.
As we will construct multiparticle Fock states,
we attach a label to the momenta, $\emo _i$,
to indicate the position of the corresponding Fock operator in
what will be {\it momentum ordered} states.
Thus, for example,
$\emo _1$ will always represent the electron with lowest momentum in the
state.
The summation indices $i,j \dots $
of the sums thus run over the
positions $1,2 \dots \enu $.
Incoming momenta $\emo _i,\pmo _i$ carry no prime
while outgoing momenta, $\emo '_i,\pmo '_i$, are primed.
As usual, `incoming' means those states on the right-hand side
of expressions.
We denote incoming and outgoing states by
$\ket[i] $ and $\bra[f] $ respectively.
For convenience
we surpress the forbidden zero momentum transfer
$\1,{0} $ terms
in the formulae.

Because of contractions between states of like-momentum
we can use the `spectator/participant' picture for the
interaction. Particles not involved in the
interaction of up to four participants
have to fulfill a pairwise momentum conservation.
This will be represented by
the symbol $\spect $
\beq
\spect = \prod_i \kd[\emo _i \emo '_i]
\prod_j \kd[\pmo _j \pmo '_j] ,
\eeq
with indices
$i,j$ running only over
spectator
positions in the Fock space vectors.

The relative sign $\rsign $
of graphs
can be calculated using reference graphs and
the number of
transpositions
of particles necessary to get to these.
This leads to terms
$\rsign = (-1) ^ {(a+b+c+d)} $ and
$\change[a,b] =
\Bigl\{
1 \r[if] b>a \, ; \;\;
0 \quad {\rm else.} \;
\Bigr\} $.

\scap{Matrix elements.}
We thus obtain the following c-number expressions for the
matrix elements of the aforementioned operators.
\beqa
& \bra[f] \res \ket[i] & =
\kd[\enu \enu ']
\prod_{n=1 \dots \enu } \kd[\emo _n \emo '_n]
\prod_{n=1 \dots \enu } \kd[\pmo _n \pmo '_n]
\sum_{n=1 \dots \enu} (\emo _n + \pmo _n) ,
\\
\nnum \\
& \bra[f] \kinE \ket[i] & =
\kd[\enu \enu ']
\prod_{n=1 \dots \enu} \kd[\emo _n \emo '_n]
\prod_{n=1 \dots \enu} \kd[\pmo _n \pmo '_n]
\sum_{n=1 \dots \enu } \1,2 \,
(\1,{\emo _n} +\1,{\pmo _n} ) ,
\\
\nnum \\
& \bra[f] \contractions \ket[i] & =
\kd[\enu \enu ']
\prod_{n=1 \dots \enu} \kd[\emo _n \emo '_n]
\prod_{n=1 \dots \enu} \kd[\pmo _n \pmo '_n] \times
\nonumber \\ &&
\sum_{n=1 \dots \enu }
\sum_{q=1 \in \resset } \1,2
\Bigl(
\1,{(\emo _n-q)^2}
+ \1,{(\pmo _n-q)^2}
- \1,{(\emo _n+q)^2}
- \1,{(\pmo _n+q)^2}  \Bigr) ,
\nnum \\
\\
& \bra[f] \seagulls \ket[i] & =
\kd[\enu \enu ']
\sum_{n,m,s,t = 1 \dots \enu }
\spect \times
\nnum \\ &&\hspace{0.5cm} \Bigr(
\kd[\emo _n + \pmo _m - \emo '_s - \pmo '_t,0]
(-1)^{( n + m + s + t)}
( \1,{(\emo _n + \pmo _m)^2} - \1,{(\emo _n - \emo '_s)^2} )
\nnum \\ &&\hspace{0.5cm}
\mbox{}+ \1,2
\kd[\emo _n + \emo _m - \emo '_s - \emo '_t,0]
(-1)^{(n+ m+ s+ t + \change[n,m] + \change[s,t] )}
\1,{(\emo _n - \emo '_s)^2}
\nnum \\ &&\hspace{0.5cm}
\mbox{}+ \1,2
\kd[\pmo _n + \pmo _m - \pmo '_s - \pmo '_t,0]
(-1)^{(n+ m+ s+ t + \change[n,m] + \change[s,t] )}
\1,{(\pmo _n - \pmo '_s)^2}
\Bigr),
\\
\nnum \\
& \bra[f] \forks \ket[i] & =
\spect
\sum_{\begin{minipage}[t]{2,4cm}
{\scriptsize \centering $n=1 \dots \enu $; \\
$m,s,t = 1 \dots \enu '$}
\end{minipage}} \kd[\enu +2,\enu '] \times
\nnum \\ &&\hspace{0.5cm}
\Bigl(
\kd[\emo _n - \emo '_m - \emo '_s - \pmo '_t,0]
(-1)^{(n+m+s+t + \change[m,s] + \enu )}
( - \1,{(\emo _n - \emo '_m)^2} )
\nonumber \\
& & \hspace{0.5cm} \mbox{}+
\kd[\pmo _n - \pmo '_m - \pmo '_s - \emo '_t,0]
(-1)^{(n+m+s+t + \change[m,s] + \enu )}
( - \1,{(\pmo '_n - \pmo '_m)^2} ) \Bigr)
\nonumber \\
& &+
\spect
\sum_{\begin{minipage}[t]{2,4cm}
\scriptsize \centering{$n,m,s=1 \dots \enu $; \\
$t = 1 \dots \enu '$}
\end{minipage}} \kd[\enu ,\enu'+2] \times
\nonumber \\ &&\hspace{0.5cm}\Bigl(
\kd[\emo _n + \emo _m + \pmo _s - \emo '_t,0]
(-1)^{(n+m+s+t + \change[n,m] + \enu')}
( - \1,{(\emo _n - \emo '_t)^2} )
\nonumber \\
& & \hspace{0.5cm}+
\kd[\pmo _n + \pmo _m + \emo _s - \pmo '_t,0]
(-1)^{( n+m+s+t+ \change[n,m] + \enu )}
( - \frac{1}{(\pmo _n - \pmo '_t)^2} ) \Bigr)
\Bigr] .
\end{eqnarray}
These analytic expressions were implemented in the
computer code we discuss in the following section.


\section{Features of the Program}
\subsection{General Run Features}
The program `ms\_main.f' was written in Fortran77 and was
compiled and run on
a Unix-system DEC 5000-260 work station,
using rou\-tines of the Numerical Algorithm Group
library NAGLIB Mark 15. It observes
the standard Fortran 77 conventions, but uses the comment sign `!'
and longer variable names which can be easily
shortened if necessary.

The `physical' input parameters are:
choice of fermion boundary conditions as a logical constant,
the harmonic resolution as a natural number,
the fermion mass and the coupling constant as positive real numbers.
The desired particle number truncation is given via two switches,
fixed particle number and maximal particle number.
The value `Zero' signifies no constraint,
so that obviously at most one of these can larger than zero,
both zero standing for no constraint at all.
Other options control the amount of output produced.

There are default input values given specifying QED(1+1) with
massive fermions and a strong coupling ($g = 2.38 m_f$).
If the inputs are internally consistent they are exchanged against the
defaults (check this, if there are problems).
A convenient deviation made
in the programming code is that the momenta and correspondingly
the harmonic resolution are stored and assigned
twice the value as in the formulae to make
them integers instead of half-integers.
For example, a harmonic resolution 10 is in fact $\res =20$
in the code and input files.

A standard test run should take about 0.2s CPU time.
Matrices of dimension of about 1500 take no more than 2000s, but use
about 10 Mbyte RAM. This is because the whole matrix
is handled as an array, the simplest possibility
for the small matrices we are dealing with.
The CPU time needed to generate the
Fock space is generally negligible
(Test Run: 0.008s) in comparison to
the time needed to calculate the LC Hamiltonian
matrix elements (0.098s) and diagonalize the matrix (0.062s). The
diagonalization time is cubic in the matrix dimension and clearly
dominates in the consumption of CPU time.

\subsection{Program structure}
The main program is `ms\_main.f'. It
\begin{itemize}
\item sets the maximum value for
the matrix dimension (= number of basis
Fock states), number of partons, resolution
and the number of lines
on one output page by including
`ms\_parameters.f'. Default values are dimension 700,
particle number 10, resolution 600 and pagelength 60 lines.
\item sets input data defaults. The values correspond to
QED(1+1) with Periodic Boundary Conditions (BC = true),
fermion mass $\ferMa =1.0$,
a rather strong coupling $\cc =2.38$
and a
small resolution $\res =20$.
Using the switch `fixed particle number =0' and
`maximal particle number =0' we include
the complete Fock space, e.g. all states with particle number up to
the highest theoretically possible one.
The output is limited
to one page for each of the Fock state basis elements, mass eigenvalues and
wave
functions.
\item reads the input data in subroutine `input' (viz. screen
output), tests
it and exchanges it against the defaults if consistent.
\item calls the master subroutine `master'.
\item  writes out to the file
`masses.out', first
a leading page with the input data
(and the defaults if these were taken for the calculation),
then the requested number of pages of Fock state basis, mass eigenvalues
and wave
functions with a header repeating the most important
input data at the top of each page.
\item as a check writes out the actual input values
to standard i/o.
\end{itemize}
In the master subroutine `master'
\begin{itemize}
\item the Fock space basis is constructed using the subroutine `con'.
\item the mass squared matrix parts $2 \res \kinE$,
$2 \res  \contractions $,
$2 \res \seagulls $
and $2 \res \forks $ are calculated using the subroutines `kinetic',
`contractions', `seagulls' and `forks'.
\item these parts are added to the correct mass squared matrix,
and this is dia\-go\-nalized using a NAGLIB routine.
\end{itemize}
The important subroutines within `con', `kinetic',
`contractions', `seagulls' and `forks' are:
\begin{itemize}
\item `fermion' -- constructs the charge singlet Fock states built
from electrons and positrons. This essentially is the heart
of the program and so we take the space here to explain
it in some detail.
This subroutine is called from the routine `con' and works as follows.
For each fixed particle number allowed,
the routine begins by building a reference state
which consists of the lowest momentum {\it electrons},
namely a sequence of integers one after the other such that
their sum does not exceed the total allowed momentum.
This is stored as the first possible electron sub-state.
 From this state further states are now generated.
The highest integer in the reference sequence is incremented.
The total momentum of the new set is checked for total
momentum saturation. If not saturated, the state is
allowed and added to
the electron sub-state array
together with the sum of the particle momenta and the particle number.
The procedure
continues with incrementation of
the highest electron momentum until total momentum
is exhausted.
The next-to-last momentum is then incremented,
all higher momenta are reset to their (lowest) starting positions,
and the routine proceeds with the highest momentum electron
in an identical way to generate more states, always checking against total
momentum saturation. Thus
the program works its way down to the lowest momentum
number. In this way an overly-large electron Fock
space is built, but with all states satisfying the Pauli
principle and momentum ordered. A positron space is
built by copying the electron states. The two spaces are then joined by
combining equal number electron and positron states.
These combined states are then
only checked against total momentum conservation
as charge conservation is already implicit.
The stored particle numbers and total momenta
of the sub-states allows
for this to be very fast, so that further improvements were
not implemented.

\item `ep\_graph' -- calculates the matrix element contributions
from electron-positron annihilation and electron-positron scattering
graphs.
It is called from the subroutine `seagulls'.
It creates all possible matrix elements for these interaction
operators by using the above-created space of states.
In a second step, matrix elements are weeded out
by checking first for particle number conservation.
Then all possible participant combinations are tested for
momentum conservation. Using the subroutine `spec',
the spectators are then checked for correct pairwise momentum conservation.
As zero momentum transfers are unphysical in our theory,
these are also excluded in a final step.
The subroutines `ee\_graph' and `pp\_graph'
are modelled on `ep\_graph' and all three taken
together generate the `seagull' matrix elements.

\item `n\_2\_sec\_e' -- calculates the matrix element contributions
from the particle number (n-2,n) sector graphs of an
incoming electron undergoing pair production.
It is called from the routine `forks'.
Again, as in `ep\_graph', all possible matrix elements are
created and then non-allowed elements removed by checking
that the number of `ingoing' and
`outgoing' particles differ by two
and that momentum is conserved in the appropriate way.
A further feature in this subroutine is the explicit inclusion
of a relative minus sign between states with two
indistinguishable electrons and/or positrons
which arises directly from the Wick expansion.
This subroutine is the template for the routines `n\_2\_sec\_p',
`n\_sec\_eep' and `n\_sec\_epp' and altogether give the
`fork' matrix elements of the light-cone Hamiltonian.

\item `spec' -- weeds out matrix elements for which spectators
do not satisfy momentum conservation. The routine
is given the participant configuration via the
general participant vector `s' holding
two ingoing and outcoming electron
and positron momenta respectively.
{}From that it is able to identify the remaining particles in a pair
of states as spectators.
For these pairwise momentum conservation is checked and when
the check fails, the states are eliminated.

\item `diag' -- handles the calling of the general black box
diagonalisation routine of the NAGLIB `F02ABF'.
We choose that routine because the QED(1+1) mass squared
operator is real and symmetric (being hermitian). It reduces
the matrix to real symmetric tridiagonal form
by Householder's
method and then uses the QL-Algorithm to extract eigenvalues
and eigenvectors. A way of checking the results is to
change this to routines `F02AGF' and `F02AXF',  which deal with general
nonsymmetric and hermitian matrixes, respectively.
The matrix is stored as an array, which causes this method to
use rather huge amounts of storage. However it is reliable,
conceptionally simple and obtains the full spectra
and wavefunctions.
Note that the eigenvectors are stored
in columns in an array called `vector'.
\end{itemize}

\subsection{Program package components}

In the package there are the following Fortran files:\\
1. ms\_main.f \\
2. ms\_parameters.f \\
3. ms\_master\_sub.f \\
4. ms\_construct\_sub.f \\
5. ms\_kinetic\_sub.f \\
6. ms\_contractions\_sub.f \\
7. ms\_seagulls\_sub.f \\
8. ms\_forks\_sub.f \\
9. ms\_spectator\_sub.f \\
10. ms\_diagonalization\_sub.f \\
11. ms\_in-out\_sub.f \\
12. ms\_input\_data \\
File 1 contains the main program,
file 2 the parameter definitions
used by all routines and
the files 3-11 the subroutines
with self-explanatory file names.
In file 12 the input of the
test run is given in a form already prepared for
automatized use.

\subsection{Reliability and Uses}
The routines shown here have already been applied to
obtain mass spectra for various coupling
constants and with different mass normalizations.
The confirmation of their reliability comes from
the very good comparison of the results with spectra
from alternate analytic and numerical methods. In
particular, these are \cite{crewther80,coleman75,bergknoff77}.
Moreover we can directly compare against the
$m_f=0$ case of the exact Schwinger model \cite{schwinger62}.
where the lowest mass eigenvalues should be $1$ and $2$.
Even a resolution of 10 ($K=20$)
already yields values of $m_1=0.99875$
and $m_2 = 1.9994$.
The wavefunctions obtained with this method
can be used to extract structure functions
\cite{elser94}. A novel example of uses for the so-obtained
spectra is the computation of thermodynamic quantities for the
given theory at finite temperature \cite{elser94a}.

A generalization of this Abelian gauge theory model
would be to include extra flavours.
Moreover these routines are already the basis for
application
to other field theories in 1+1 dimensions
such as QED with massive photons \cite{martinovic94}.

\subsection{Test Run}
In Appendix (\ref{screen}) we give a sample
screen output with the sequence of data inputting,
checking and confirmation. The input values used here
are the default values.
The data file output of a test corresponding to this
default input is shown in Appendix (\ref{data}).
This should be the basis of checking a test run of the
program package.
As a second option, in Appendix (\ref{input})
we give a suggested input file
which can be piped into the program while
running our routines in batch mode.


\section{Concluding Remarks}

In this paper we have described a program to solve
gauge theories for the bound state spectra and
wavefunctions via the formal method of Discretized
Light-Cone Quantization. The concrete test case for which we
gave details for was quantum electrodynamics in one
space and one time dimension. This model theory
shares many features with the real problem
to which DLCQ is eventually intended to be applied,
namely quantum chromodynamics. These common features
are the presence of Lorentz and gauge symmetry, fermions and
a linear confining potential. However, the absence of
spin, color and transverse momentum
quantum numbers allow a simple exposition of the
essentials of the programming procedure. We thus descibed routines
which constructed a Fock space of electrons and
positrons, and which then computed matrix elements
of the Hamiltonian for charge singlet
combinations of the Fock states. Diagonalization of this
Hamiltonian has been used to generate, for example, spectra of
bound states of electron-positron pairs in two-dimensions.

Extension to non-Abelian gauge theories, namely
QCD in two dimensions, is technically
more complicated in that color statistics must be
introduced into the Fock space construction and more
complicated interactions built into the code.
However these represent a `grafting' onto the type of code
we explicitly discussed here.
Further extension to include more space-dimensions is
also feasible in the same spirit pursued here,
where color is compounded by the presence of genuinely
dynamical gluons in the Fock space construction.
In both cases, the methods used here have been the basis
for codes developed in, for example, \cite{woelz,heyssler94}.
These two examples far from exhaust the variations of
simplifications on full QCD which could be constructed
in order to build toward the full problem and for which
DLCQ, as a general method, could be insightful. As
a further example we mention the novel application to
two-dimensional Yang-Mills theory coupled to scalar adjoint matter
by \cite{demeterfi94}.
For a treatment of full QCD, the major problems which could
change some aspects of the programming method we describe are the
renormalization and zero mode problems. These difficult formal problems
are under active consideration.

\begin{appendix}
\newpage

\section{Screen output}
\label{screen}

\renewcommand{\baselinestretch}{0.7}

\vspace{-0.5cm}
\begin{scriptsize}
\hspace{-1cm}
\begin{verbatim}
  1
+-------------------------------------------------------------------------+
  2	    |
  |
  3	    |         The Output of       'ms'       on your terminal screen
  |
  4	    |
  |
  5
+-------------------------------------------------------------------------+
  6	    |
  |
  7	    |         please give input data:
  |
  8	    |
  |
  9	    |         Next input data set? (1=yes)
  |
 10	    |
  |
 11	    |         boundary conditions
  |
 12	    |         (antiperiodic = T, periodic = F)
  |
 13	    |         boundary conditions are                              T
  |
 14	    |
  |
 15	    |         harmonic resolution
  |
 16	    |         harmonic resolution is                              20
  |
 17	    |
  |
 18	    |         fermion mass (with decimal point)
  |
 19	    |         fermion mass is                                   1.00
  |
 20	    |
  |
 21	    |         coupling constant (with decimal point)
  |
 22	    |         coupling constant is                              2.38
  |
 23	    |
  |
 24	    |         particle number limits
  |
 25	    |          (one of these must be zero;
  |
 26	    |           both zero is a full fock space request)
  |
 27	    |
  |
 28	    |         fixed particle number
  |
 29	    |         fixed particle number is                             0
  |
 30	    |
  |
 31	    |         maximal particle number
  |
 32	    |         maximal particle number is                           0
  |
 33	    |
  |
 34	    |         front page (1=yes)
  |
 35	    |         front page printed
  |
 36	    |
  |
 37	    |         maximal number of Fock space pages
  |
 38	    |         maximal number of Fock space pages is                1
  |
 39	    |
  |
 40	    |         maximal number of eigenvalue pages
  |
 41	    |         maximal number of eigenvalue pages is                1
  |
 42	    |
  |
 43	    |         maximal number of wavefunction pages
  |
 44	    |         maximal number of wavefunction pages is              1
  |
 45	    |
  |
 46
+-------------------------------------------------------------------------+
 47	    |
  |
 48	    |         check output - you calculated with data:
  |
 49	    |
  |
 50	    |         BC - boundary conditions:               True   (antiperiodic)
  |
 51	    |
  |
 52	    |         res - harmonic resolution:                20
  |
 53	    |
  |
 54	    |         m_f - fermion mass:                     1.00
  |
 55	    |
  |
 56	    |         c_c - coupling constant:                2.38
  |
 57	    |
  |
 58	    |         fixed particle number:                     0
  |
 59	    |
  |
 60	    |         maximal particle number:                   6
  |
 61	    |
  |
 62	    |         Output is:       1 of max    2 page(s) of Fock   space
  |
 63	    |                          1 of max    1 page(s) of mass  values
  |
 64	    |                          1 of max   12 page(s) of eigenvectors
  |
 65	    |
  |
 66	    |         CPU-time for the DLCQ matrix.        Dimension is   42 .
  |
 67	    |         total time:      0.168 s    fock space:          0.008 s
  |
 68	    |         matrix:          0.098 s    diagonalisation:     0.062 s
  |
 69	    |
  |
 70
+-------------------------------------------------------------------------+
 71
 72
+-------------------------------------------------------------------------+
 73	    |
  |
 74	    |         The Output of       'ms'       on your terminal screen
  |
 75	    |
  |
 76
+-------------------------------------------------------------------------+
 77	    |
  |
 78	    |         please give input data:
  |
 79	    |
  |
 80	    |         Next input data set? (1=yes)
  |
\end{verbatim}
\end{scriptsize}

\renewcommand{\baselinestretch}{1.25}
\section{Data output}
\label{data}

\renewcommand{\baselinestretch}{0.69}

\vspace{-0.5cm}
\begin{scriptsize}
\hspace{-1cm}
\begin{verbatim}
   1	    'ms' - output                                            18:07:06
15-Jul-94
   2	    BC: True         m_f= 1.00        c_c= 2.38         Nmax=  6
res= 20
   3
   4
+-------------------------------------------------------------------------+
   5	    |
   |
   6	    |                                  'ms'
   |
   7	    |
   |
   8
+-------------------------------------------------------------------------+
   9	    |
   |
  10	    |         program 'ms' calculated the mass spectrum of QED 1+1
   |
  11	    |         with the accepted input data:
   |
  12	    |
   |
  13	    |         BC - boundary conditions:               True
(antiperiodic)   |
  14	    |
   |
  15	    |         res - harmonic resolution:                20
   |
  16	    |
   |
  17	    |         m_f - fermion mass:                     1.00
   |
  18	    |
   |
  19	    |         c_c - coupling constant:                2.38
   |
  20	    |
   |
  21	    |         fixed particle number:                     0
   |
  22	    |
   |
  23	    |         maximal particle number:                   0
   |
  24	    |         full Fock space - max. part. number:       6
   |
  25	    |
   |
  26	    |         Output is:       1 of max    2 page(s) of Fock   space
   |
  27	    |                          1 of max    1 page(s) of mass  values
   |
  28	    |                          1 of max   12 page(s) of eigenvectors
   |
  29	    |
   |
  30	    |         CPU-time for the DLCQ matrix.        Dimension is   42 .
   |
  31	    |         total time:      0.168 s    fock space:          0.008 s
   |
  32	    |         matrix:          0.098 s    diagonalisation:     0.062 s
   |
  33	    |
   |
  34
+-------------------------------------------------------------------------+

::::::::::: page 2 ::::::::::

  61	    'ms' - output                                            18:07:04
05-Jul-94
  62	    BC: True         m_f= 1.00        c_c= 2.38         Nmax=  6
res= 20
  63
  64
+-------------------------------------------------------------------------+
  65	    |  Fock space basis used by 'ms'                         1 -   42 of
42 |
  66
+-------------------------------------------------------------------------+
  67
  68	    state mass^2  part    lowest 5 electron momenta   lowest 5 positron
momenta
  69
  70	        1  21.05     2                            1
  19
  71	        2   7.84     2                            3
  17
  72	        3   5.33     2                            5
  15
  73	        4   4.40     2                            7
  13
  74	        5   4.04     2                            9
  11
  75	        6   4.04     2                           11
   9
  76	        7   4.40     2                           13
   7
  77	        8   5.33     2                           15
   5
  78	        9   7.84     2                           17
   3
  79	       10  21.05     2                           19
   1
  80	       11  48.00     4                       3    1                     15
   1
  81	       12  34.87     4                       3    1                     13
   3
  82	       13  32.48     4                       3    1                     11
   5
  83	       14  31.75     4                       3    1                      9
   7
  84	       15  45.54     4                       5    1                     13
   1
  85	       16  32.48     4                       5    1                     11
   3
  86	       17  30.22     4                       5    1                      9
   5
  87	       18  44.68     4                       7    1                     11
   1
  88	       19  31.75     4                       7    1                      9
   3
  89	       20  29.71     4                       7    1                      7
   5
  90	       21  44.44     4                       9    1                      9
   1
  91	       22  31.75     4                       9    1                      7
   3
  92	       23  44.68     4                      11    1                      7
   1
  93	       24  32.48     4                      11    1                      5
   3
  94	       25  45.54     4                      13    1                      5
   1
  95	       26  48.00     4                      15    1                      3
   1
  96	       27  32.48     4                       5    3                     11
   1
  97	       28  19.56     4                       5    3                      9
   3
  98	       29  17.52     4                       5    3                      7
   5
  99	       30  31.75     4                       7    3                      9
   1
 100	       31  19.05     4                       7    3                      7
   3
 101	       32  31.75     4                       9    3                      7
   1
 102	       33  19.56     4                       9    3                      5
   3
 103	       34  32.48     4                      11    3                      5
   1
 104	       35  34.87     4                      13    3                      3
   1
 105	       36  29.71     4                       7    5                      7
   1
 106	       37  17.52     4                       7    5                      5
   3
 107	       38  30.22     4                       9    5                      5
   1
 108	       39  32.48     4                      11    5                      3
   1
 109	       40  31.75     4                       9    7                      3
   1
 110	       41  60.19     6                  5    3    1                 7    3
   1
 111	       42  60.19     6                  7    3    1                 5    3
   1

::::::::::: page 3 ::::::::::

 121	    'ms' - output                                            18:07:04
05-Jul-94
 122	    BC: True         m_f= 1.00        c_c= 2.38         Nmax=  6
res= 20
 123
 124
+-------------------------------------------------------------------------+
 125	    |  Mass^2 eigenvalues calculated by 'ms'                 1 -   42 of
42 |
 126
+-------------------------------------------------------------------------+
 127
 128	       1:  3.2217  7.0677  11.2069  11.7232  12.3816  14.4305  16.0938
17.5337
 129	       9: 17.7096 17.9037  19.7839  20.0023  21.6207  21.7420  21.8782
22.8458
 130	      17: 22.9126 23.7122  23.9960  25.2102  25.7435  26.1450  26.6680
27.8514
 131	      25: 28.9435 29.4743  33.4945  33.6035  35.1072  35.3365  37.4794
38.3057
 132	      33: 41.4285 43.1637  46.1262  46.9805  49.7538  51.6405  57.6122
59.4818
 133	      41: 66.6438 75.3417

::::::::::: page 4 ::::::::::

 181	    'ms' - output                                            18:07:04
05-Jul-94
 182	    BC: True         m_f= 1.00        c_c= 2.38         Nmax=  6
res= 20
 183
 184
+-------------------------------------------------------------------------+
 185	    |  'ms' wavefunction (times 10)                           page   1 of
 1 |
 186
+-------------------------------------------------------------------------+
 187
 188	    state       1       2        3        4        5        6        7
   8
 189	    mass^2   3.22    7.07    11.21    11.72    12.38    14.43    16.09
17.53
 190
 191	       1:  1.9568 -2.9886  -3.3115   0.9786  -1.6094   2.6508  -2.7312
1.6289
 192	       2:  2.7551 -3.9533  -3.6808   0.9977  -1.5350   1.8788  -1.4141
0.4102
 193	       3:  3.2898 -3.8657  -1.6452   0.1565   0.0947  -2.2260   2.7494
-2.2542
 194	       4:  3.6375 -2.8099   1.6692  -1.1390   2.4319  -1.4712   2.6845
0.6297
 195	       5:  3.8105 -1.0277   4.1583  -1.7331   0.4649  -0.6044  -2.4939
0.1992
 196	       6:  3.8105  1.0277   4.1583   1.7331  -0.4649   0.6044  -2.4939
0.1992
 197	       7:  3.6375  2.8099   1.6692   1.1390  -2.4319   1.4712   2.6845
0.6297
 198	       8:  3.2898  3.8657  -1.6452  -0.1565  -0.0947   2.2260   2.7494
-2.2542
 199	       9:  2.7551  3.9533  -3.6808  -0.9977   1.5350  -1.8788  -1.4141
0.4102
 200	      10:  1.9568  2.9886  -3.3115  -0.9786   1.6094  -2.6508  -2.7312
1.6289
 201	      11:  0.0508  0.1227   0.5743  -0.2382   0.4703  -1.4469   1.1210
-0.7516
 202	      12:  0.0107  0.1923   0.4471  -0.3046   0.7164   2.2283  -0.6816
1.8602
 203	      13:  0.0012  0.1634   0.1320  -0.1184  -2.1230  -1.0377  -0.9995
-2.6779
 204	      14: -0.0002  0.0651  -0.0075   2.4650   1.2108  -0.1620   1.0982
3.8333
 205	      15:  0.0484  0.3630   0.9904  -0.5019   1.0885   0.9985  -0.1918
1.5882
 206	      16:  0.0083  0.3870   0.4591  -0.4231  -1.6197   1.4899  -2.0400
-0.8839
 207	      17: -0.0004  0.2494   0.0161   2.6933  -1.0611  -1.3940   0.5794
0.9497
 208	      18:  0.0275  0.5813   0.7688  -0.4837  -1.2287   0.3499  -2.1919
-0.5414
 209	      19: -0.0003  0.4728   0.0473   2.4318  -0.3335   1.4770   0.1299
0.9056
 210	      20: -0.0021  0.1830  -0.1102  -0.1730  -2.6526  -1.3541   0.8304
-0.7469
 211	      21:  0.0000  0.6695   0.0000   2.0339  -0.1441   0.3881   0.0000
0.0000
 212	      22: -0.0092  0.3973  -0.3850  -0.4348  -1.7352   1.6020   1.7598
-0.0200
 213	      23: -0.0275  0.5813  -0.7688  -0.4837  -1.2287   0.3499   2.1919
0.5414
 214	      24: -0.0123  0.2058  -0.4299  -0.3354   0.8182   2.7150   0.6096
-1.5719
 215	      25: -0.0484  0.3630  -0.9904  -0.5019   1.0885   0.9985   0.1918
-1.5882
 216	      26: -0.0508  0.1227  -0.5743  -0.2382   0.4703  -1.4469  -1.1210
0.7516
 217	      27:  0.0123  0.2058   0.4299  -0.3354   0.8182   2.7150  -0.6096
1.5719
 218	      28:  0.0021  0.2121   0.1534  -0.1939  -3.1555  -1.6260  -1.3350
-1.7284
 219	      29: -0.0001  0.0895  -0.0056   3.8877   1.9939  -0.2040   0.5719
2.1043
 220	      30:  0.0092  0.3973   0.3850  -0.4348  -1.7352   1.6020  -1.7598
0.0200
 221	      31:  0.0000  0.3037   0.0000   3.4922  -1.4084  -1.8976   0.0000
0.0000
 222	      32:  0.0003  0.4728  -0.0473   2.4318  -0.3335   1.4770  -0.1299
-0.9056
 223	      33: -0.0021  0.2121  -0.1534  -0.1939  -3.1555  -1.6260   1.3350
1.7284
 224	      34: -0.0083  0.3870  -0.4591  -0.4231  -1.6197   1.4899   2.0400
0.8839
 225	      35: -0.0107  0.1923  -0.4471  -0.3046   0.7164   2.2283   0.6816
-1.8602
 226	      36:  0.0021  0.1830   0.1102  -0.1730  -2.6526  -1.3541  -0.8304
0.7469
 227	      37:  0.0001  0.0895   0.0056   3.8877   1.9939  -0.2040  -0.5719
-2.1043
 228	      38:  0.0004  0.2494  -0.0161   2.6933  -1.0611  -1.3940  -0.5794
-0.9497
 229	      39: -0.0012  0.1634  -0.1320  -0.1184  -2.1230  -1.0377   0.9995
2.6779
 230	      40:  0.0002  0.0651   0.0075   2.4650   1.2108  -0.1620  -1.0982
-3.8333
 231	      41:  0.0000 -0.0001   0.0045   0.0106   0.0180   0.0109  -0.0823
-0.0831
 232	      42:  0.0000  0.0001   0.0045  -0.0106  -0.0180  -0.0109  -0.0823
-0.0831
 233
\end{verbatim}
\end{scriptsize}

\renewcommand{\baselinestretch}{1.25}

\section{Batch job input data}
\label{input}

\renewcommand{\baselinestretch}{0.7}

\begin{scriptsize}
\hspace{-1cm}
\begin{verbatim}
  1	    1                 !---------- input file for 'ms' -----------
  2	       .true.            !BC          boundary condition
  3	       20                !res         harmonic resolution
  4	       1.0               !m_f         fermion mass
  5	       2.38              !g           coupling constant
  6	       0                 !N_fix       fixed   parton number
  7	       0                 !N_max       maximal parton number
  8	       1                 !pages(1)    front page   print option
  9	       1                 !pages(2)    Fock space   print option
 10	       1                 !pages(3)    eigenvalue   print option
 11	       1                 !pages(4)    wavefunction print option
 12	   -1                 !---------- input file for 'ms' -----------
\end{verbatim}
\end{scriptsize}

\renewcommand{\baselinestretch}{1.25}
\end{appendix}

\newpage


\begin{thebibliography}{ABC}

\bibitem{kogut79}
J.B. Kogut,
Rev.Mod.Phys. {\bf 51} (1979) 659.

\bibitem{butler93}
F. Butler, H. Chen, J. Sexton, A. Vaccarino, D. Weingarten,
Phys.Rev.Lett. {\bf 70} (1993) 2849.

\bibitem{dirac49}
P.A.M. Dirac, Rev.Mod.Phys. {\bf 21} (1949) 392.

\bibitem{overview}
S.J. Brodsky, H.C. Pauli, in
{\it Recent Aspects of Quantum Fields}, eds. H. Mitter, H. Gausterer
(Lecture Notes in Physics, Vol. 396, Springer-Verlag, Heidelberg 1991);
S.J. Brodsky, G. McCartor, H.C. Pauli, S.S. Pinsky,
Particle World {\bf 3} (1993) 109.

\bibitem{woelz}
F. W\"olz,
{\it On the Spectrum of Normal modes in Quantum Chromodynamics
and the Theory of the Effective Interaction from the
Tamm-Dancoff Method}, PhD thesis,
Ruprecht-Karls-Universit\"at, Heidelberg, 1995.


\bibitem{hiller93}
J.J. Wivoda, J.R. Hiller,
Phys.Rev. {\bf D48} (1993) 4647.

\bibitem{vandesande94}
S.S. Pinsky, B. van de Sande,
Phys.Rev. {\bf D49} (1994) 2001; \\
S.S. Pinsky, B. van de Sande, J.R. Hiller,
Phys.Rev. {\bf D51} (1995) 726.

\bibitem{demeterfi94}
K. Demeterfi, I.R. Klebanov, G. Bhanot,
Nucl.Phys. {\bf B418} (1994) 15.

\bibitem{tam94}
A. Tam, C.J. Hamer, C.M. Yung,
{\it Light-Cone Quantization Approach to Quantum
Electrodynamics in (2+1) Dimensions},
New South Wales University Preprint PRINT-94-0182 (1994).

\bibitem{epb}
Th. Eller, H.C. Pauli, S.J. Brodsky, Phys.Rev. {\bf D35}
(1987) 1493; \\
Th. Eller, H.C. Pauli, Z.Phys. {\bf C42} (1989) 59.

\bibitem{perry}
R.J. Perry, K.G. Wilson,
Nucl.Phys. {\bf B 403} (1993) 587.

\bibitem{kalloniatis}
A.C. Kalloniatis, H.C. Pauli, Z.Phys. {\bf C 63} (1994) 161 \\
A.C. Kalloniatis, D.G. Robertson, Phys.Rev. {\bf D 50} (1994) 5262;\\
A.C. Kalloniatis, H.-C. Pauli, S.S. Pinsky,
Phys.Rev. {\bf D 50} (1994) 6633; \\
H.-C. Pauli, A.C. Kalloniatis, S.S. Pinsky, {\it Towards Solving
QCD -- the Transverse Zero Modes in Light-Cone Quantization},
MPI f\"ur Kernphysik Heidelberg preprint
MPIH-V3-1995 (MPIH-V23-1994) (Revised).
To appear in Phys.Rev. D.
\bibitem{kogut70}
J. Kogut, D. Soper, Phys.Rev. {\bf D1} (1970) 2901.

\bibitem{pauli85}
H.C. Pauli, S.J. Brodsky, Phys.Rev. {\bf D32} (1985) 1993 and 2001.

\bibitem{elser94}
S. Elser, {\it The
Spectrum of ${\rm QED}_{1+1}$ in the framework
of the DLCQ method}, diploma thesis,
Ruprecht-Karls-Universit\"at Heidelberg, 1994;
{\it The Spectrum of ${\rm QED}_{1+1}$ in the framework
of the DLCQ method}, in {\it Hadron Structure '94 (Proceedings)}, eds.
J. Urb\'an and J. Vrl\'akov\'a,
Kosice, 1994.

\bibitem{lanczos}
C. Lanczos, J.Res.Nat.Bur.Stand. {\bf 45} (1950) 255

\bibitem{crewther80}
D.P. Crewther, C.J. Hamer, Nucl.Phys. {\bf B170}
(1980) 353.

\bibitem{coleman75}
S. Coleman, R. Jackiw, L. Susskind, Ann.Phys.(N.Y.)
{\bf 93} (1975) 267; \\
S. Coleman, Ann.Phys.(N.Y.)
{\bf 101} (1976) 239.


\bibitem{bergknoff77}
H. Bergknoff, Nucl.Phys. {\bf B122} (1977) 215.

\bibitem{schwinger62}
J. Schwinger, Phys.Rev. {\bf 128} (1962) 2425.

\bibitem{elser94a}
S. Elser, A.C. Kalloniatis, in preparation.

\bibitem{martinovic94}
L. Martinovic, S. Elser, work in progress.

\bibitem{heyssler94}
M. Heyssler, {\it Numerical Methods for Solving ${\rm QCD}_{1+1}$ in
the context of the DLCQ Method}, diploma thesis,
Ruprecht-Karls-Universit\"at Heidelberg, 1994;
{\it Constituent Quark Picture out of QCD in two-dimensions ---
on the Light-Cone},
MPI f\"ur Kernphysik Heidelberg preprint MPIH-V25-1994 (Revised).
Submitted to Phys.Lett. B.



\end{thebibliography}
\end{document}